\documentclass[ALICE,manyauthors]{cernphprep}
\newif\ifcernpp
\cernpptrue

\usepackage{hyperref}
\usepackage{lineno}

\usepackage{graphicx}
\usepackage[loose]{units}
\usepackage{upgreek}
\usepackage{color}
\usepackage[usenames,dvipsnames]{xcolor}
\usepackage{amsmath,amssymb,amsbsy} 
\ifcernpp
\usepackage{cite}
\usepackage[comma,square,numbers,sort&compress]{natbib}
\else
\fi

\ifcernpp
\else
\fi

\begin{document}%

\newcommand{\PbPb}{\textnormal{Pb--Pb}}
\newcommand{\AuAu}{\textnormal{Au--Au}}
\newcommand{\pp}{\ensuremath{\mbox{p}\mbox{p}}}
\newcommand{\pt}{\ensuremath{p_\mathrm{T}}}
\newcommand{\pT}{\pt}
\newcommand{\ptch}{\ensuremath{p_\mathrm{T,\,ch\;jet}}}
\newcommand{\deltaptch}{\ensuremath{\delta p_\mathrm{T,\,ch}}}
\newcommand {\snnbf}{\ensuremath{\mathbf{\sqrt{s_{_{\mathrm{NN}}}} }}}
\newcommand {\snn}{\ensuremath{\sqrt{s_{_{\mathrm{NN}}}} }}

\newcommand{\CKBNOTE}[1]{{\bf CKB:  #1}} 
\newcommand{\MWNOTE}[1]{{\bf MW:  #1}} 
\newcommand{\DPNOTE}[1]{{\bf DP:  #1}} 
\newcommand{\FBNOTE}[1]{{\bf FB:  #1}} 
\newcommand{\KRNOTE}[1]{{\bf KR:  #1}} 

\newcommand {\SlopeCentral} {\ensuremath{(297 \pm 12^\mathrm{stat}\pm 41^\mathrm{syst})\,\mathrm{MeV}}}
\newcommand {\SlopeCentralNoSubtr} {\ensuremath{(304 \pm 11^\mathrm{stat}\pm 40^\mathrm{syst})\,\mathrm{MeV}}}
\newcommand {\SlopeSemiCentral} {\ensuremath{(410 \pm 84^\mathrm{stat} \pm 140^\mathrm{syst})\,\mathrm{MeV}}}
\newcommand {\SlopeSemiCentralNoSubtr} {\ensuremath{(407 \pm 61^\mathrm{stat} \pm 96^\mathrm{syst})\,\mathrm{MeV}}}
\newcommand {\FitrangeCentral} {\ensuremath{0.9 < \pT < \unit[2.1]{GeV}/c}}
\newcommand {\FitrangeSemiCentral} {\ensuremath{1.1 < \pT < \unit[2.1]{GeV}/c}}

\ifcernpp
\begin{titlepage}
\PHyear{2015}
\PHnumber{254}      
\PHdate{14 September}  
\else
\journal{Physics Letters B}
\fi
%
%
\title{Direct photon production in Pb-Pb collisions at $\mathbf{\sqrt{s_\mathrm{NN}} = 2.76\,TeV}$}
\ifcernpp
\ShortTitle{Direct photon production in Pb-Pb}   
%
\Collaboration{ALICE Collaboration\thanks{See Appendix~\ref{app:collab} for the list of collaboration members}}
\ShortAuthor{ALICE Collaboration} 
\else
\author{ALICE Collaboration}
\fi
\begin{abstract}
Direct photon production at mid-rapidity in Pb-Pb collisions at $\snn = \unit[2.76]{TeV}$ was studied in the transverse momentum range $0.9 < \pT < \unit[14]{GeV}/c$. Photons were detected with the highly segmented electromagnetic calorimeter PHOS and via conversions in the ALICE detector material with the $e^+e^-$ pair reconstructed in the central tracking system. The results of the two methods were combined and direct photon spectra were measured for the 0--20\%, 20--40\%, and 40--80\% centrality classes. For all three classes, agreement was found with perturbative QCD calculations for $\pT \gtrsim \unit[5]{GeV}/c$. Direct photon spectra down to $\pT \approx \unit[1]{GeV}/c$ could be extracted for the 20--40\% and 0--20\% centrality classes. The significance of the direct photon signal for $\FitrangeCentral$ is $2.6\sigma$ for the 0--20\% class. The spectrum in this $\pt$ range and centrality class can be described by an exponential with an inverse slope parameter of $\SlopeCentral$. State-of-the-art models for photon production in heavy-ion collisions agree with the data within uncertainties.
\end{abstract}

\ifcernpp
\end{titlepage}
\setcounter{page}{2}
\else
\maketitle
\fi

\section{Introduction}

The theory of the strong interaction, Quantum ChromoDynamics (QCD), predicts a transition from ordinary nuclear matter to a state where quarks and gluons are no longer confined to hadrons \cite{Borsanyi:2013bia,Bazavov:2011nk}. The creation and study of this deconfined partonic state, the Quark-Gluon Plasma (QGP), is the major objective in the experimental program of heavy-ion collisions at the Relativistic Heavy Ion Collider (RHIC) \cite{Adams:2005dq,Adcox:2004mh,Arsene:2004fa,Back:2004je} and the Large Hadron Collider (LHC) \cite{Aamodt:2010pb,Aamodt:2010cz,ATLAS:2011ag,Chatrchyan:2011pb,Aamodt:2010pa,ATLAS:2011ah,Abelev:2012hxa,Aad:2010bu,Chatrchyan:2011sx}.

Direct photons, defined as photons not originating from hadron decays, are a valuable tool to study details of the evolution of the medium created in heavy-ion collisions.  Unlike hadrons, direct photons are produced at all stages of the collision and escape from the hot nuclear matter basically unaffected \cite{Thoma:1994fd}, delivering direct information on the conditions at the time of production: {\it prompt} direct photons produced in hard scatterings of incoming partons provide information on parton distributions in nuclei; deconfined quark-gluon matter as well as hadronic matter created in the course of the collision emit {\it thermal} direct photons, carrying information about the temperature, collective flow and space-time evolution of the medium \cite{Gale:2009gc}. Different transverse momentum ($\pT$) regions are dominated by photons emitted at different stages of the collision. Prompt direct photons follow a power law spectrum and dominate at high transverse momentum ($\pT \gtrsim \unit[5]{GeV}/c$). At lower transverse momenta ($\pT \lesssim \unit[4]{GeV}/c$) one expects contributions from the thermalized partonic and hadronic phases with an approximately exponential spectrum \cite{Kapusta:1991qp,Turbide:2003si}. In addition, other direct photon production mechanisms, like the interaction of hard scattered partons with the medium ("jet-photon conversion") \cite{Fries:2002kt,Turbide:2007mi}, may be important for $\pT \lesssim \unit[10]{GeV}/c$.

The direct photon spectrum at low $\pT$, therefore, contains information on the initial temperature and space-time evolution of the thermalized medium created in heavy-ion collisions. The observed thermal direct photon spectrum is a sum of contributions from all stages of the collision after thermalization, where the earliest, hottest stage and later, cooler stages can make comparable contributions \cite{Shen:2013vja}. High photon emission rates at the largest temperatures in the early stage are compensated by an expanded space-time volume and blue-shift due to radial flow in the later stage. This complicates the interpretation of inverse slope parameters of direct photon spectra, but a correlation between the slope and the initial temperature still exists \cite{d'Enterria:2005vz}.

The first measurement of a direct photon spectrum in relativistic A-A collisions was presented by the WA98 collaboration \cite{Aggarwal:2000th}. The direct photon yield was measured at the CERN SPS in central Pb-Pb collisions at $\snn = \unit[17.3]{GeV}$ in the range $1.5< \pT< \unit[4]{GeV}/c$. The signal can be interpreted either as thermal photon radiation from a quark-gluon plasma and hadronic gas or as the effect of multiple soft scatterings of the incoming partons without the formation of a QGP \cite{Turbide:2003si}. The PHENIX experiment measured the direct photon spectrum in Au-Au collisions at $\snn = \unit[200]{GeV}$ in the range $1 \lesssim \pT \lesssim \unit[20]{GeV}/c$  \cite{Adler:2005ig,Afanasiev:2012dg}. It was found that at high $\pT$ ($5 \lesssim \pT \lesssim \unit[21]{GeV}/c$) the direct photon spectrum measured in Au-Au collisions agrees with the one measured in \pp\ collisions at the same energy after scaling with the number of binary nucleon-nucleon collisions ($N_\mathrm{coll}$). Scaling of high-$\pT$ direct photon production with $N_\mathrm{coll}$ in Pb-Pb collisions at LHC energy was confirmed by the ATLAS \cite{Aad:2015lcb} and CMS \cite{Chatrchyan:2012vq} experiments in the measurement of {\it isolated} photons, i.e., photons with little hadronic energy in a cone around them, in the ranges  $22< \pT < \unit[280]{GeV}/c$ and $20<\pT < \unit[80]{GeV}/c$, respectively. The absence of suppression of high $p_T$ isolated photons in A-A collisions with respect to $N_\mathrm{coll}$ scaled pp collisions, in contrast to the observed suppression of hadrons, is consistent with the latter being due to energy loss of hard scattered quarks and gluons in the medium.

Direct photon production at low $\pT$ ($\lesssim \unit[3]{GeV}/c$) in Au-Au collisions at $\snn = \unit[200]{GeV}$ was studied by the PHENIX experiment in the measurement of virtual photons ($e^+e^-$ pairs from internal conversions) \cite{Adare:2008ab} and with real photons \cite{Adare:2014fwh}. A clear excess of direct photons above the expectation from scaled \pp\ collisions was observed. The excess was parameterized by an exponential function with inverse slope parameters $\unit[T_{\rm eff} = 221 \pm 19^{\rm stat} \pm 19^{\rm syst}]{MeV}$ (virtual photon method \cite{Adare:2008ab}) and $\unit[T_{\rm eff} = 239 \pm 25^{\rm stat} \pm 7^{\rm syst}]{MeV}$ (real photon method \cite{Adare:2014fwh}) for the 0--20\% most central collisions. The measured spectrum can be described by models assuming thermal photon emission from hydrodynamically expanding hot matter with initial temperatures in the range $\unit[300-600]{MeV}$ \cite{Adare:2009qk}. The measurement of a direct-photon azimuthal anisotropy (elliptic flow), which was found to be similar in magnitude to the pion elliptic flow at low $\pT$ in Au-Au collisions at $\snn = \unit[200]{GeV}$ \cite{Adare:2011zr}, provides a further important constraint for models. The simultaneous description of the spectra and elliptic flow of direct photons currently poses a challenge for hydrodynamic models \cite{Chatterjee:2013naa}. 

In this letter, the first measurement of direct photon production for $\pT\lesssim \unit[14]{GeV}/c$ in Pb-Pb collisions at $\snn = \unit[2.76]{TeV}$ is presented.

\section{Detector setup}
Photons were measured using two independent methods: by the Photon Conversion Method (PCM) and with the electromagnetic calorimeter PHOS. In the conversion method, the electron and positron tracks from a photon conversion were measured with the Inner Tracking System (ITS) and/or the Time Projection Chamber (TPC).

The ITS \cite{Aamodt:2008zz} consists of two layers of Silicon Pixel Detectors (SPD) positioned at a radial distance of \unit[3.9]{cm} and \unit[7.6]{cm}, two layers of Silicon Drift Detectors (SDD) at \unit[15.0]{cm} and \unit[23.9]{cm}, and two layers of Silicon Strip Detectors (SSD) at \unit[38.0]{cm} and \unit[43.0]{cm}. The two innermost layers cover a pseudorapidity range of $|\eta|<2$ and $|\eta| < 1.4$, respectively. The TPC \cite{Alme:2010ke} is a large (85~m$^3$) cylindrical drift detector filled with a Ne-CO$_2$-N$_2$ (90-10-5) gas mixture. It covers the pseudorapidity range $|\eta|<0.9$ over the full azimuthal angle with a maximum track length of 159 reconstructed space points. With the magnetic field of $B=\unit[0.5]{T}$, $e^+$ and $e^-$ tracks can be reconstructed down to $\pT \approx \unit[50]{MeV}/c$, depending on the position of the conversion point. The TPC provides particle identification via the measurement of the specific energy loss (d$E$/d$x$) with a resolution of 5.2\% in pp collisions and 6.5\% in central Pb-Pb collisions \cite{Abelev:2014ffa}. The ITS and the TPC were aligned with respect to each other to the level of less than $\unit[100]{\mu m}$ using cosmic-ray and \pp\ collision data \cite{Aamodt:2010aa}. Particle identification is furthermore provided by the Time-of-Flight (TOF) detector \cite{Akindinov:2009zze} located at a radial distance of $370 < r < \unit[399]{cm}$. This detector consists of Multigap Resistive Plate Chambers (MRPC) and provides timing information with an intrinsic resolution of \unit[50]{ps}.

PHOS \cite{Dellacasa:1999kd} is an electromagnetic calorimeter which consists of three modules installed at a distance of \unit[4.6]{m} from the interaction point. It subtends $260^\circ<\varphi<320^\circ$ in azimuth and $|\eta|<0.13$ in pseudorapidity. Each module consists of 3584 detector cells arranged in a matrix of $64\times 56$ lead tungstate crystals each of size \unit[$2.2\times 2.2 \times 18$]{cm$^3$}. The signal from each cell is measured by an avalanche photodiode (APD) associated with a low-noise charge-sensitive preamplifier. To increase the light yield, reduce electronic noise, and improve energy resolution, the crystals, APDs, and preamplifiers are cooled to a temperature of $-25~^\circ$C. The resulting energy resolution is $\sigma_{E}/E=(1.3\%/E) \oplus (3.3\%/\sqrt{E})\oplus 1.12$\%, where $E$ is in GeV. The PHOS channels were calibrated in \pp\ collisions by aligning the $\pi^0$ peak position in the two-photon invariant mass distribution.

Two scintillator hodoscopes (V0-A and V0-C) \cite{Cortese:2004aa} subtending $2.8 < \eta < 5.1$ and $-3.7 < \eta < -1.7$, respectively, were used in the minimum bias trigger in the \PbPb\ run. The sum of the amplitudes of V0-A and V0-C served as a measure of centrality in the \PbPb\ collisions.

\section{Data analysis}
This analysis is based on data recorded by the ALICE experiment in the first LHC heavy-ion run in the fall of 2010. The detector readout was triggered by the minimum bias interaction trigger based on trigger signals from the V0-A, V0-C, and SPD detectors. The efficiency for triggering on a \PbPb\ hadronic interaction ranged between 98.4\% and 99.7\%, depending on the minimum bias trigger configuration.  The events were divided into centrality classes according to the V0-A and V0-C summed amplitudes. Only events in the centrality range 0--80\% were used in this analysis. To ensure a uniform track acceptance in pseudorapidity $\eta$, only events with a primary vertex within $\pm \unit[10]{cm}$ from the nominal interaction point along the beam line ($z$-direction) were used. After offline event selection, $13.6 \times 10^6$ events were available for the PCM analysis and $17.7 \times 10^6$ events for the PHOS analysis. 

\begin{figure}[h]
\unitlength\textwidth
\centering
\includegraphics[width=0.7\linewidth]{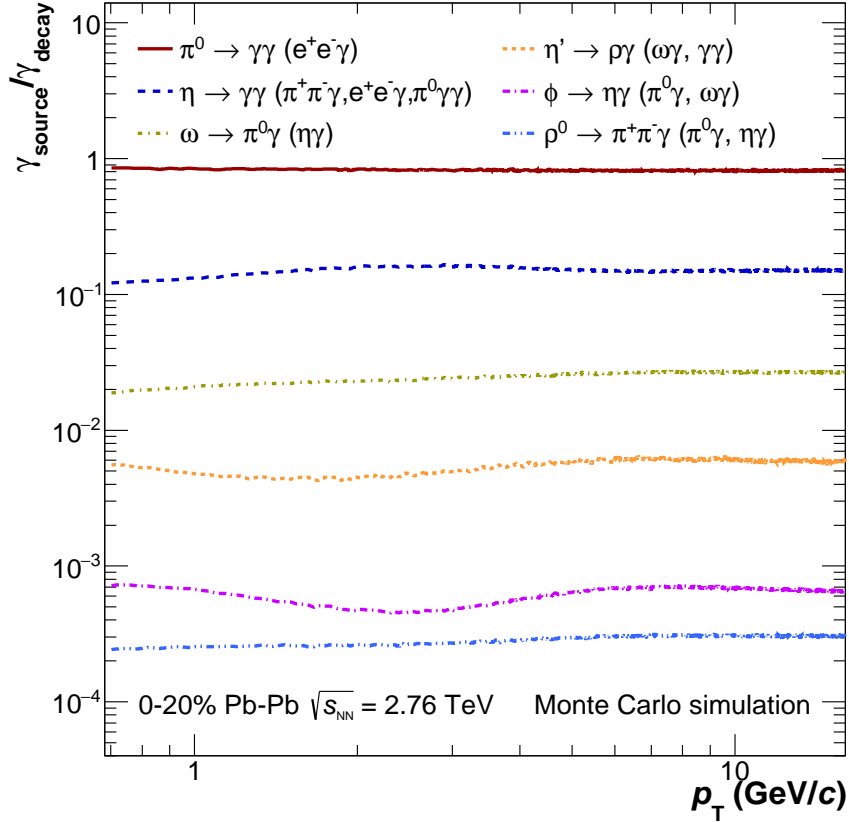}
\caption{
\label{fig:Cocktail} 
(Color online) Relative contributions of different hadrons to the total decay photon spectrum as a function of the decay photon transverse momentum (PCM case). 
}
\end{figure}

The direct photon yield is extracted on a statistical basis from the inclusive photon spectrum by comparing the measured photon spectrum to the spectrum of photons from hadron decays. The yield of $\pi^0$s, which contribute about $80-85$\% of the decay photons (cf. Fig.~\ref{fig:Cocktail}), was measured simultaneously with the inclusive photon yield. Besides photons from $\pi^0$ decays, the second and third most important contributions to the decay photon spectrum come from $\eta$ and $\omega$ decays. 

An excess of direct photons above the decay photon spectrum can be quantified by the $\pT$ dependent double ratio
\begin{equation}
  \label{eq:doubleratio}
  R_\gamma  \equiv \left . \frac{\gamma_{\mathrm{incl}}}{\pi^0_{\mathrm{param}}} \right / \frac{\gamma_{\mathrm{decay}}}{\pi^0_{\mathrm{param}}}
 = \frac{\gamma_{\mathrm{incl}}}{\gamma_{\mathrm{decay}}}, 
\end{equation}
where $\gamma_{\mathrm{incl}}$ is the measured inclusive photon spectrum, $\pi^0_{\mathrm{param}}$ a parameterization of the measured $\pi^0$ spectrum, and $\gamma_{\mathrm{decay}}$ the calculated decay photon spectrum. The PCM and PHOS $\pi^0$ measurements are described in \cite{Abelev:2014ypa}. The double ratio has the advantage that some of the largest systematic uncertainties cancel partially or completely. Using the double ratio, the direct photon yield  can be calculated from the inclusive photon yield as
\begin{equation}\label{eq:subs}
  \gamma_{\mathrm{direct}} = \gamma_{\mathrm{incl}}-\gamma_{\mathrm{decay}} = (1- \frac{1}{R_\gamma})\cdot \gamma_{\mathrm{incl}}.
\end{equation}
The PCM and PHOS analyses were performed independently. Combined direct photon spectra were determined based on combined double ratios and combined inclusive photon spectra. In contrast to taking the average of the PCM and PHOS direct-photon spectra, this approach allowed us to use the information from both measurements also when one measurement of $R_\gamma$ fluctuated below unity.

In the PCM analysis, photons are reconstructed via a secondary vertex finding algorithm which provides displaced vertices with two opposite-charge daughters. The positively and negatively charged daughter tracks are required to contain reconstructed clusters in the TPC. Only tracks with a transverse momentum above \unit[50]{MeV}/$c$ and a ratio of the number of reconstructed TPC clusters over the number of findable TPC clusters (accounting for track length, spatial location and momentum) larger than 0.6 were considered. To identify $e^+$ and $e^-$, the specific energy loss in the TPC \cite{Abelev:2014ffa} was required to be within a band of $[-3\sigma , 5\sigma ]$ around the average electron $\mathrm{d}E/\mathrm{d}x$, and be more than $3\sigma$ above the average pion $\mathrm{d}E/\mathrm{d}x$ (where the second condition is only applied for tracks with $\pT > \unit[0.4]{GeV}/c$). Tracks with an associated signal in the TOF detector were only accepted as electron candidates if they were consistent with the electron hypothesis within a $\pm 5\sigma$ band. The vertex finding algorithm uses the Kalman filter technique for the decay/conversion point and four momentum determination of the neutral parent particle ($V^0$) \cite{Abelev:2012cn}. $V^0$s result from $\gamma$ conversions but also from strange particle decays ($K_s^0$, $\Lambda$ or $\bar\Lambda$). Further selection was performed on the level of the reconstructed $V^0$. $V^0$s with a decay point with radius $r<\unit[5]{cm}$ were rejected to remove $\pi^0$ and $\eta$ Dalitz decays. The transverse momentum component $q_T = p_e \sin \theta_{V0,e}$ \cite{podolanski:1954iii} of the electron momentum, $p_e$, with respect to the $V^0$ momentum was restricted to $q_T < \unit[0.05]{GeV}/c$. Based on the invariant mass of the $e^{+} e^{-}$ pair and the pointing of the $V^0$ to the primary vertex, the vertex finder calculates a $\chi^2(\gamma)$ value which reflects the level of consistency with the hypothesis that the $V^0$ comes from a photon originating from the primary vertex. A selection based  on this  $\chi^2(\gamma)$ value was used to further reduce contamination in the photon sample. Random associations of electrons and positrons were further reduced by making use of the small opening angle of the $e^+e^-$ pair from photon conversions at the conversion point.

The raw photon spectrum, constructed from the secondary vertex candidates passing the selection described above, was corrected for the reconstruction efficiency, the acceptance and the contamination. The detector response was simulated for $\PbPb$ collisions using HIJING \cite{Wang:1991hta} together with the GEANT~3.21 transport code \cite{Brun:1987ma}. The resulting efficiency correction is dominated by the conversion probability of photons in the ALICE material. The integrated material budget of the beam pipe, the ITS and the TPC for $r<\unit[1.8]{m}$ corresponds to $(11.4 \pm 0.5)\%$ of a radiation length $X_0$, resulting in a photon conversion probability that saturates at about 8.5\% for $\pT \gtrsim \unit[2]{GeV}/c$ \cite{Abelev:2012cn,Abelev:2014ffa}. The photon finding efficiency for converted photons is of the order of 50--65\% over the measured $\pT$ range for all centralities. The purity of the photon candidate sample for $\pT < \unit[3]{GeV}/c$ extracted from simulation is 98--99\% in peripheral and 91--97\% in the most central collisions. Furthermore, secondary photon candidates, mainly photons from the decay $K_s^0\rightarrow 2\pi^0\rightarrow 4\gamma$, not removed by the $\chi^2(\gamma)$ selection, were subtracted statistically based on the measured $K_{s}^0$ spectrum \cite{Abelev:2013xaa}. A correction of less than 2\% for photons from pile-up collisions was applied for the 40--80\% class for $\pt < \unit[2]{GeV}/c$. At higher $\pt$ and for more central classes this correction is negligible.

\begin{table}[ht]
\centering
\begin{tabular}{lcccccc}
     \hline
Centrality     & \multicolumn{2}{c}{0--20\%} & \multicolumn{2}{c}{20--40\%}  & \multicolumn{2}{c}{40--80\%}\\
\hline
$\pT$ (GeV/$c$)               &  1.2 & 5.0 &  1.2 & 5.0 &  1.2 & 5.0 \\
\hline
{\bf $\pmb{\gamma_\mathrm{incl}}$ yield}  \\
Track quality (A)    		& 0.6 & 0.6 & 0.2 & 0.2 & 0.2 & 0.7 \\
Electron PID    (A,B) 	        & 1.5 & 6.9 & 0.9 & 4.8 & 0.7 & 4.0 \\
Photon selection  (A,B)       & 4.0 & 1.8 & 2.4 & 2.1 & 1.5 & 1.3 \\
Material (C) 			& 4.5 & 4.5 & 4.5 & 4.5 & 4.5 & 4.5 \\
$\pmb{\gamma_\mathrm{incl}/\pi^0}$ \\
Track quality (A)   		& 0.7 & 1.7 & 0.8 & 0.4 & 0.6 & 1.3 \\
Electron PID (A,B)    	& 1.2 & 4.8 & 0.9 & 3.8 & 0.9 & 4.0 \\
Photon selection (A,B) 	& 3.2 & 3.2 & 3.0 & 1.5 & 2.5 & 2.4 \\
$\pi^0$ yield (A)   		& 1.6 & 2.9 & 1.7 & 2.7 & 0.5 & 3.0 \\
Material (C)    			& 4.5 & 4.5 & 4.5 & 4.5 & 4.5 & 4.5 \\
{\bf $\pmb{\gamma_\mathrm{decay}/\pi^0}$} \\
$\pi^0$ spectrum (B) 	& 0.5 & 1.2 & 0.8 & 1.8 & 0.5 & 3.2 \\
$\eta$ yield (C)  		& 1.4 & 1.4 & 1.4 & 1.4 & 1.4 & 1.4 \\
$\eta$ shape (B)  		& 1.6 & 0.5 & 1.2 & 0.2 & 1.0 & 0.2 \\
\hline
Total $R_\gamma$     		& 6.2 & 8.1 & 5.7 & 7.0 & 5.7 & 8.3 \\
Total $\gamma_\mathrm{incl}$      
					& 6.2 & 8.5 & 5.2 & 6.9 & 4.8 & 6.2 \\
\end{tabular} 
\caption{\label{tab:sys_inc_PCM}
Summary of the systematic uncertainties of the PCM analysis in percentage. Uncertainties are characterized according to three categories: point-by-point uncorrelated (A), correlated in $\pT $ with magnitude of the relative uncertainty varying point-by-point (B), and constant fractional uncertainty (C). Items in the table with categories (A,B) summarize sources of uncertainties which are either of type~A or B.}

\end{table}

In the PHOS analysis, clusters (each cell of the cluster must have at least one common edge with another cell of the cluster) were used as photon candidates. To estimate the photon energy, the energies of cells with centers within a radius $R_{\rm core} = \unit[3.5]{cm}$ from the cluster center of gravity were summed. Compared to the full cluster energy, this {\it core energy} ($E_\mathrm{core}$) is less sensitive to overlaps with low-energy clusters in a high multiplicity environment. The non-linearity in the conversion of the reconstructed to the true photon energy introduced by this approach is reproduced by GEANT3 Monte Carlo simulations. The contribution of hadronic clusters was reduced by requiring $E_{\rm cluster}>\unit[0.3]{GeV}$, $N_{\mathrm{cells}}>2$ and by accepting only clusters above a minimum lateral cluster dispersion \cite{Abelev:2014ypa}. The latter selection rejects hadrons punching though the crystal and producing a large signal in the photodiode of a single cell. 
With a minimum time between bunch crossings of \unit[525]{ns}, possible pile-up contributions from other bunch crossings is removed by a loose cut on the cluster arrival time $|t|<\unit[150]{ns}$. For systematic uncertainty studies, photons were also reconstructed with a $\pT$-dependent dispersion cut and with a charged particle veto (CPV) cut on the distance between the PHOS cluster position and the position of extrapolated charged tracks on the PHOS surface to suppress clusters from charged particles \cite{Abelev:2014ypa}. Both dispersion and CPV cuts were tuned using \pp\ collision data to provide a photon efficiency at the level of 96--99\%. 

\begin{table}
\centering
\begin{tabular}{lcccccc}
     \hline
Centrality     & \multicolumn{2}{c}{0--20\%} & \multicolumn{2}{c}{20--40\%}  & \multicolumn{2}{c}{40--80\%}\\
\hline
$\pT$ (GeV/$c$) &  2 & 10 &  2 & 10 &  2 & 10 \\
\hline
{\bf $\pmb{\gamma_\mathrm{incl}}$ yield}  \\
Efficiency (B)   &  3.0  & 3.0 & 0.7 & 0.7 & 2.5  &  2.5 \\
Contamination (B) &  2.0  & 2.0 & 1.3 & 1.3 & 2.9  & 0.5  \\
Conversion (C)   & 1.7   & 1.7 & 1.7 & 1.7 & 1.7  & 1.7 \\
Acceptance (C)   & 1.0   & 1.0 & 1.0 & 1.0 & 1.0  &  1.0 \\
$^*$Global E scale (B) & 9.6 & 9.0 & 6.1 & 5.9 & 5.8  & 6.3 \\
$^*$Non-linearity (B) & 2.2  & 0.1 & 2.1 & 0.1 & 2.0  & 0.1 \\
$\pmb{\pi^0}$ {\bf  yield} \\
Yield extraction (A) & 2.7  & 4.0 & 3.1 & 5.2 & 1.8 & 2.9 \\
Efficiency  (B)  & 1.8  & 1.8  & 2.7 & 2.2 & 2.5 & 2.5 \\ 
Acceptance  (C)  & 1.0   & 1.0  & 1.0 & 1.0 & 1.0 & 1.0 \\
Pileup      (C)  & 1.0   & 1.0 & 1.0 & 1.0& 1.0 & 1.0 \\
Feed-down   (B)  & 2.0   & 2.0 & 2.0 & 2.0 & 2.0 & 2.0 \\
\multicolumn{6}{l}{\bf $\pmb{\gamma_\mathrm{decay}/\pi^0}$}\\
$\pi^0$ spectrum (B)  & 1.3 &  4.3 & 1.8 & 1.8 & 1.8 & 1.8 \\
$\eta$ contribution (B) & 2.2 & 1.7 & 2.2& 1.6 & 2.1 & 1.6 \\
\hline
Total $R_\gamma$  &6.8 & 7.9  &5.9 & 6.5 &6.1  & 6.0 \\
Total $\gamma_\mathrm{incl}$  &12.4 &12.7 & 9.7&10.0 &9.8  & 9.6 \\
\end{tabular} 
\caption{\label{tab:sys_inc_PHOS}
Summary of systematic uncertainties of the PHOS analysis in percentage.  Uncertainties are characterized according to three categories: point-by-point uncorrelated (A), correlated in $\pT $ with magnitude of the relative uncertainty varying point-by-point (B), and constant fractional uncertainty (C). Uncertainties marked with * cancel in the double ratio $R_\gamma$. }

\end{table}

The product of acceptance and efficiency ($A\cdot\varepsilon$) was estimated by embedding simulated photon clusters into real events and applying the standard reconstruction. PHOS properties (energy and position resolutions, residual de-calibration, absolute calibration, non-linear energy response) were tuned in the simulation to reproduce the $\pT$ dependence of the $\pi^0$ peak position and width \cite{Abelev:2014ypa}. In peripheral events, $A\cdot\varepsilon$ for the default selection (no dispersion cut, no CPV cut) has a value of about 0.022 at $\pT = \unit[1]{GeV}/c$. For higher $\pT$, $A\cdot\varepsilon$ decreases and saturates at about 0.018 for $\pT \gtrsim \unit[5]{GeV}/c$. The decrease of $A\cdot\varepsilon$ with $\pT$ results from the use of $E_\mathrm{core}$. In central collisions, $A\cdot\varepsilon$ increases by up to about $10$\% due to cluster overlaps. Applying the dispersion and CPV cuts, the efficiency is reduced by 5--10\% in peripheral collisions and the centrality dependence becomes negligible.

The contamination of the photon spectrum measured with PHOS originates mainly from $\pi^{\pm}$ and $\bar p$, $\bar n$ annihilation in PHOS, with other contributions being much smaller. Application of the dispersion and CPV cuts reduces the overall contamination at $\pT \approx 1.5 $ GeV/$c$ from about 15\% to 2--3\% and down to 1--2\% at $\pT\sim 3$--$\unit[4]{GeV}/c$. The subtraction of contamination is based on a data driven approach: the probability to pass the CPV and dispersion cuts and the calorimeter response to hadrons are estimated using identified $\pi^\pm$, $\bar p$ tracks; the photon candidate spectra, measured with different cuts (default, dispersion, CPV, both) were decomposed into $\gamma$, $\pi^\pm$, $\bar p$ and $\bar n$ contributions, assuming equal contamination from $\bar p$ and $\bar n$. The contamination calculated in this way agrees with that estimated from a HIJING simulation. Finally, the photon contribution from $K_s^0\to 2\pi^0\to4\gamma$ decays was subtracted based on the measured $K_{s}^0$ spectrum \cite{Abelev:2013xaa} as in the PCM analysis.

To calculate the $\gamma_\mathrm{decay}/\pi^0$ ratio, a Monte Carlo approach was used to simulate particle decays into photons both for the PCM and the PHOS analysis. The largest contributions come from $\pi^0$, $\eta$, and $\omega$ decays. Contributions of other hadrons were also included but were found to be negligible. To allow for a cancellation of some uncertainties common to the photon and $\pi^0$ yield in Eq.~(\ref{eq:doubleratio}), each analysis (PCM, PHOS) used the $\pi^0$ spectrum measured with the respective method.

The $\eta$ meson contribution is estimated by using two approaches which assume: 
(i) transverse mass ($m_{\mathrm{T}}$) scaling of the $\pi^0$ and the $\eta$ spectrum which is consistent with measurements at RHIC \cite{Adler:2006bv,Adare:2009qk} or 
(ii) that the $\pT$ spectrum of the $\eta$ has the same shape as the $K^0_s$ spectrum \cite{Abelev:2013xaa} as both particles should be affected by radial flow in the same way due to their similar masses. The maximum deviation between these two cases occurs at $\pt \approx \unit[2.5]{GeV}/c$ where (i) corresponds to a $\eta/\pi^0$ ratio of about 0.4 whereas (ii) gives a ratio of about 0.5. The absolute yield of $\eta$ mesons in both cases was fixed at $\pT > \unit[5]{GeV}/c$ to reproduce the measured $\eta/\pi^0$ ratio at $\snn = \unit[200]{GeV}$: $0.46\pm 0.05$ \cite{Adare:2012wg}. The statistical precision of the $\eta$ signal in the 2010 and 2011 data sets is too low to further constrain these two assumptions with a measurement of the $\eta$ spectrum. The average of these two cases is used for the decay photon calculation, while half the difference is taken as a contribution to the systematic uncertainty of the $\eta$ meson contribution in addition to the normalization uncertainty quoted above. The contribution of $\omega$ meson decay photons is below $\sim 3\%$ and $m_{\mathrm{T}}$ scaling of the measured $\pi^0$ spectrum with $(\mathrm{d}N_\mathrm{\omega}/\mathrm{d}m_\mathrm{T})/(\mathrm{d}N_\mathrm{\pi^0}/\mathrm{d}m_\mathrm{T}) = 0.9$ is used \cite{Adare:2011ht}. The relative contributions of the different hadrons to the total decay photon spectrum are shown in Fig.~\ref{fig:Cocktail}.

The main sources of systematic uncertainties in the determination of the inclusive photon spectrum and $R_\gamma$ for the PCM analysis are listed in Table~\ref{tab:sys_inc_PCM}. The two largest uncertainties are related to the material budget of the ALICE detector and the Monte Carlo-based efficiency corrections to the inclusive photon and $\pi^0$ spectra. The material budget uncertainty was estimated in pp collisions by comparing the measured number of converted photons (normalized to the measured charged particle multiplicity) with GEANT simulation results in which particle yields from PYTHIA and PHOJET were used as input. Uncertainties related to track selection and electron identification were estimated by variation of the cuts. For instance, we observe a small variation in results depending on the minimum threshold for electron tracks. This is most likely related to different tracking performance for real data and in the Monte Carlo simulation for low-$\pT$ particles ($\pT \lesssim \unit[50]{MeV}/c$). The uncertainty related to the choice of this threshold was estimated by increasing the minimum $\pT$ from $\unit[50]{MeV}/c$ up to $\unit[100]{MeV}/c$. Uncertainties related to falsely reconstructed electron-positron pairs from Dalitz decays as conversion pairs were obtained by varying the minimum radial distance $R_\mathrm{min}$ of reconstructed electron tracks from the standard value of $R_\mathrm{min} = \unit[5]{cm}$ up to  $R_\mathrm{min} = \unit[10]{cm}$. The estimation of the systematic uncertainty of the electron selection includes a contribution estimated by the variation of the d$E$/d$x$ cuts.
 
In the double ratio $R_\gamma$, many uncertainties partially cancel. The uncertainties on $R_\gamma$ were therefore obtained by evaluating the effect of cut variations directly on $R_\gamma$. Uncertainties related to the decay photon spectrum are similar for PCM and PHOS analyses: they include the uncertainty due to the $\pi^0$ spectrum parameterization, difference of shapes of $\pi^0$ spectra measured by PCM and PHOS, and uncertainties due to the shape and absolute normalization of the $\eta$ spectrum. Uncertainties due to contributions of other hadrons are negligible. 

The main systematic uncertainties of the PHOS analysis are summarized in Table~\ref{tab:sys_inc_PHOS}. For the inclusive photon spectrum, the uncertainty of the efficiency calculation is estimated comparing the PID cut efficiency in Monte Carlo and real data. The contamination uncertainty is estimated comparing the photon purity calculated with a data driven approach and with Monte Carlo HIJING simulations. The conversion probability is estimated comparing $\pi^0$ yields in pp collisions with and without magnetic field. The global energy and non-linearity uncertainties, which mostly cancel in $R_\gamma$, are estimated comparing calibrations based on the $\pi^0$ peak position and on the electron $E/p$ peak position. The centrality dependence of the energy scale uncertainty results from the larger background under the $\pi^0$ peak in central events and therefore larger uncertainties in the peak position.

\begin{figure}[h]
\unitlength\textwidth
\centering
\includegraphics[width=0.7\linewidth]{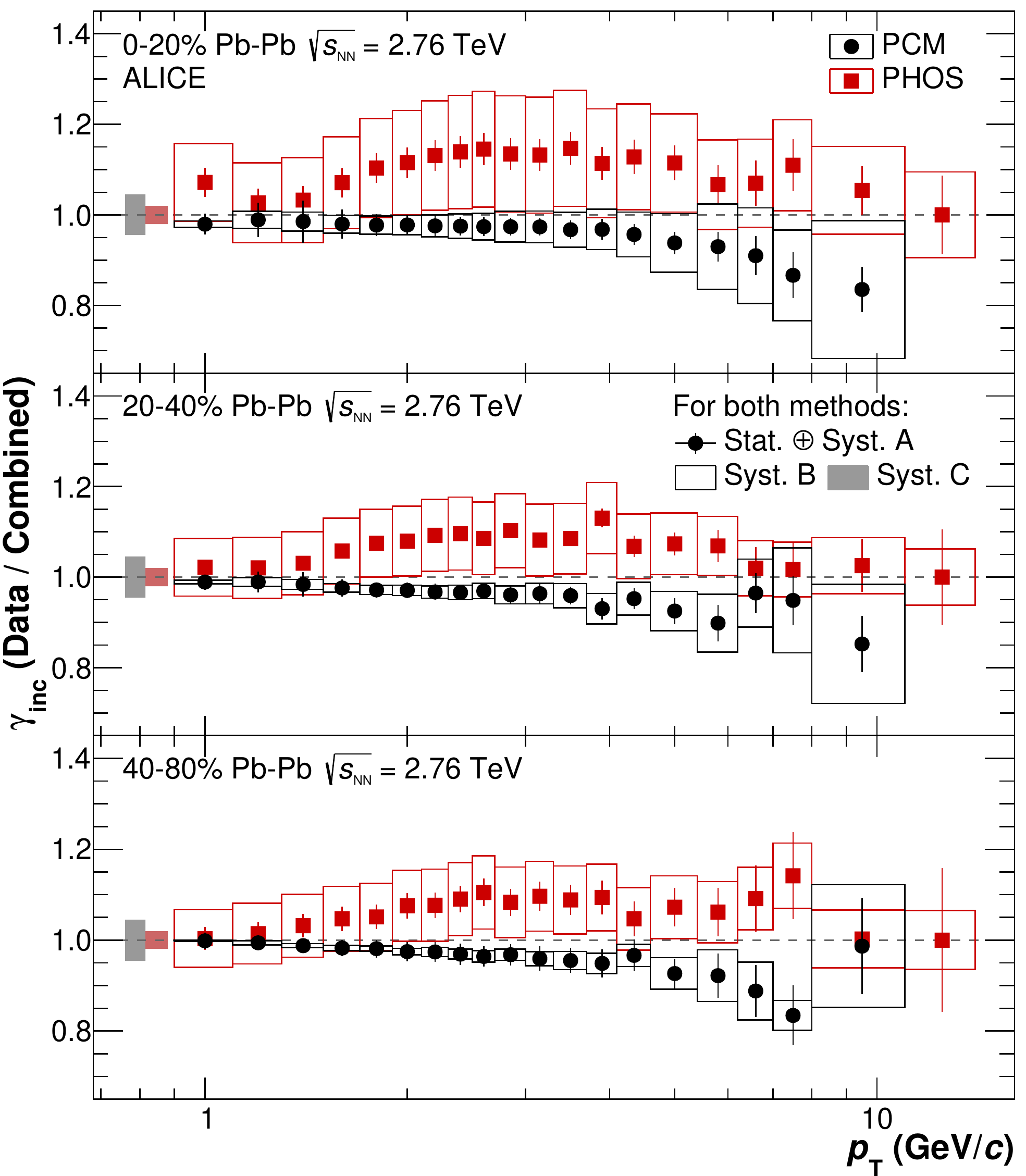}
\caption{
\label{fig:Incspectra} 
(Color online) Comparison of inclusive photon spectra measured with PCM and PHOS in the 0--20\%, 20--40\%, and 40--80\% centrality classes. The individual spectra were divided by the corresponding combined PCM and PHOS spectrum. The shown errors only reflect the uncertainties of the individual measurements. The boxes around unity indicate normalization uncertainties (type~C).
}
\end{figure}

A more detailed description of the single photon selection and especially of the additional $\pi^0$ uncertainties for both the PCM and PHOS analyses can be found in Ref.~\cite{Abelev:2014ypa}. 

The comparison of the individual PHOS and PCM inclusive photon spectra, normalized to the averaged spectrum, is shown in Fig.~\ref{fig:Incspectra}. Statistical and point-to-point uncorrelated systematic uncertainties (type~A) are combined and presented as error bars, point-to-point correlated systematic uncertainties (type~B) are shown as boxes, and common normalization systematic uncertainties (type~C) are shown as bands around unity. The uncertainties are dominated by $\pT$-correlated contributions. 
\begin{figure}[h]
\unitlength\textwidth
\centering
\includegraphics[width=0.7\linewidth]{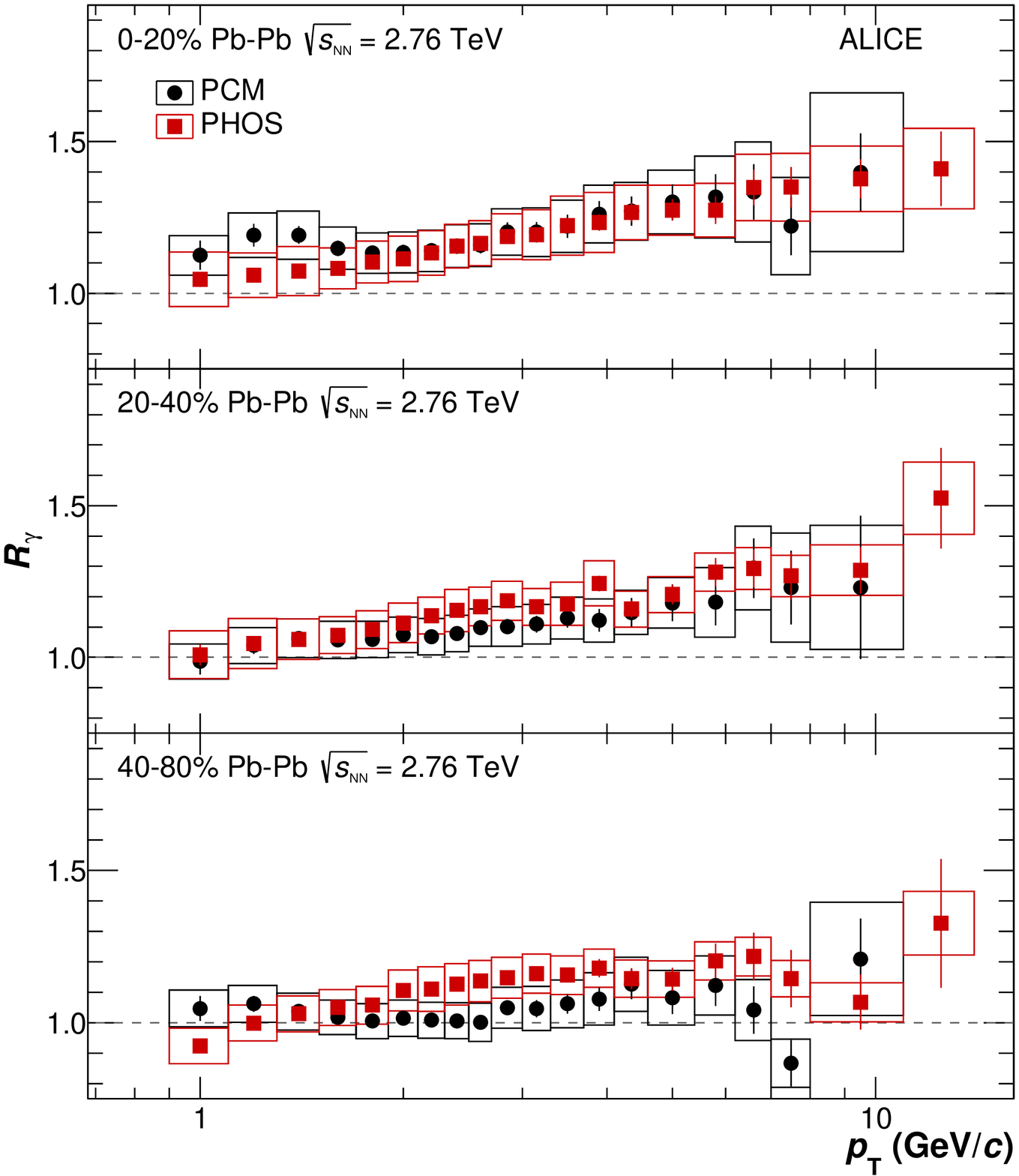}
\caption{\label{fig:compDR} 
(Color online) Comparison of double ratios $R_\gamma$ measured with PCM and PHOS for the 0--20\%, 20--40\%, and 40--80\%  centrality classes. Error bars reflect the statistical and type A systematic uncertainty, the boxes represent the type~B and C systematic uncertainties. The cancellation of uncertainties (energy scale, material budget) in the double ratio $R_\gamma$ is taken into account in the shown systematic uncertainties.}
\end{figure}
The individual PHOS and PCM double ratios are shown in Fig.~\ref{fig:compDR}. The partial cancellation of the energy scale uncertainties (PHOS) and the material budget uncertainties (PCM) is taken into account in the shown uncertainties.

The level of agreement between the PHOS and PCM inclusive photon spectra and double ratios $R_\gamma$ was quantified taking into account the correlation of the uncertainties in $\pt$ and centrality. To this end, pseudo data points for the ratio of the PHOS and PCM inclusive photon spectra and double ratios were generated simultaneously for all three centrality classes under the assumption of the null hypothesis that the ratio is unity for all points, i.e., that both measurements result from the same original distribution. The type~B and C systematic uncertainties give rise to a shifted baseline, around which the pseudo data points are drawn from a Gaussian with a standard deviation given by the statistical and type~A uncertainties. A test statistic $t$ was defined as the sum of the squared differences of the pseudo data points with respect to the null hypothesis in units of the type~A and statistical uncertainties. A $p$-value was calculated as the fraction of pseudo experiments with values of $t$ larger than observed in the real data \cite{Sinervo:2002sa}. The corresponding significance in units of the standard deviation of a one-dimensional normal distribution was calculated based on a two-tailed test. The PHOS and PCM inclusive photon spectra were found to agree within 1.2 standard deviations, the PHOS and PCM double ratios agree within 0.4 standard deviations. 

\section{Results}

\begin{figure}[h]
\unitlength\textwidth
\centering
\includegraphics[width=0.7\linewidth]{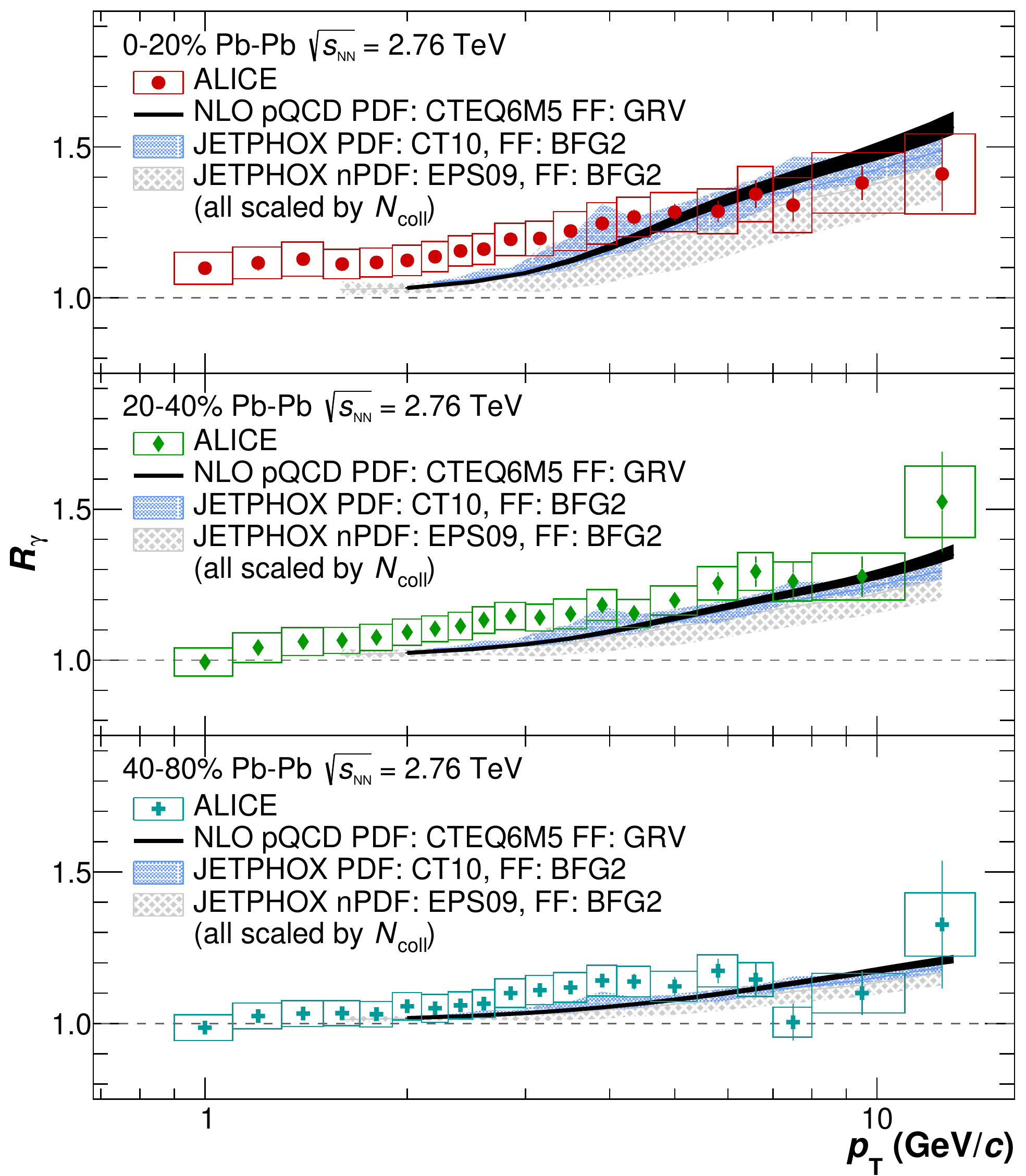}
\caption{\label{fig:rgamma} 
(Color online) Combined PCM and PHOS double ratio $R_\gamma$ in the 0--20\%, 20--40\%, and 40--80\% centrality classes compared with pQCD calculations for nucleon-nucleon collisions scaled by the number of binary collisions for the corresponding Pb-Pb centrality class. The dark blue curve is a calculation from Refs.~\cite{Gordon:1993qc,Vog97a} which uses the GRV photon fragmentation function \cite{Gluck:1992zx}. The JETPHOX calculations \cite{Klasen:2013mga} were performed with two different parton distribution functions, CT10 \cite{Lai:2010vv} and EPS09 \cite{Eskola:2009uj}, and the BFG II fragmentation function \cite{Bourhis:1997yu}.}
\end{figure}

\begin{figure}[h]
\unitlength\textwidth
\centering
\includegraphics[width=0.7\linewidth]{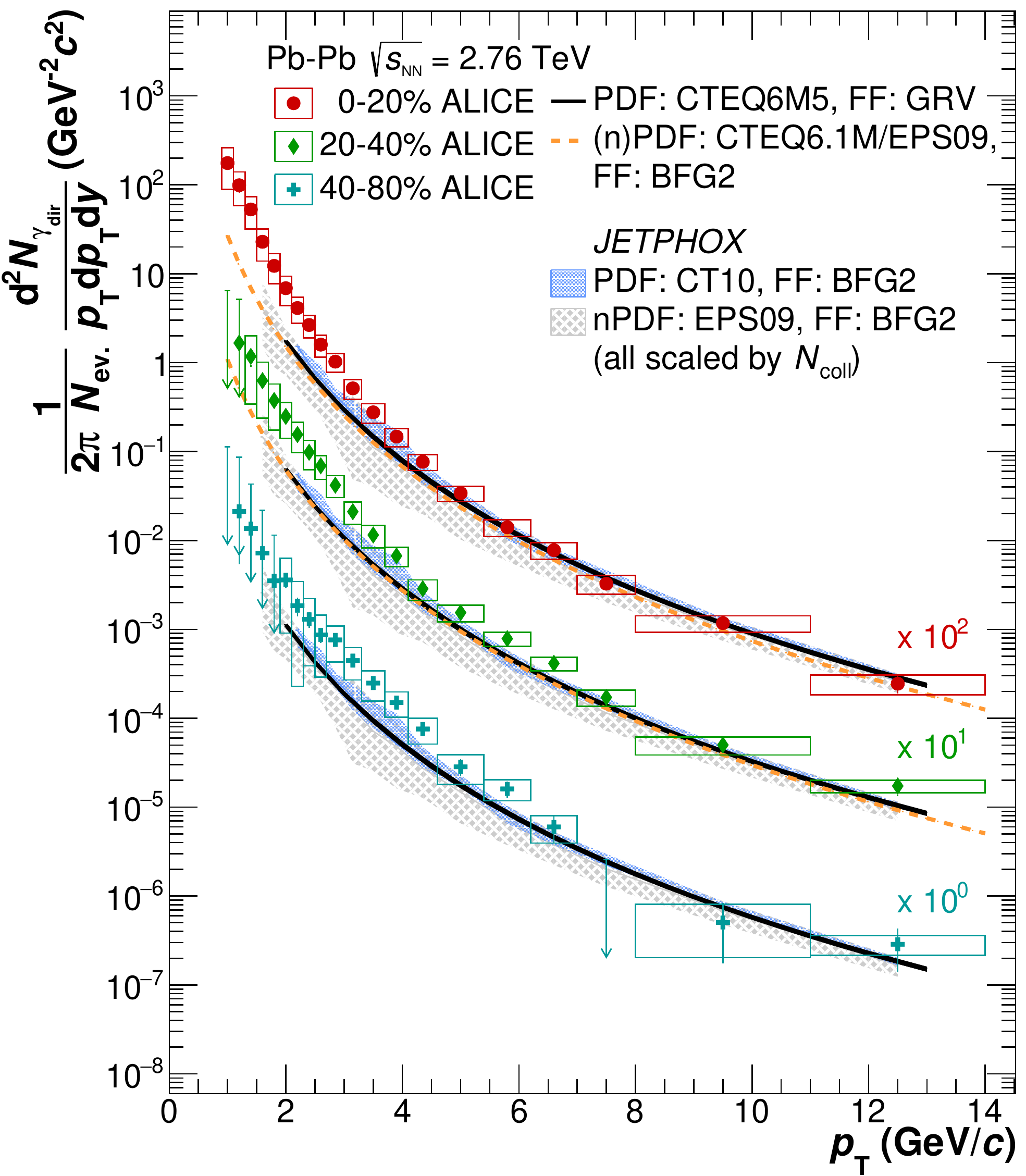}
\caption{
\label{fig:spectra} 
(Color online) Direct photon spectra in \PbPb\ collisions at $\snn=\unit[2.76]{TeV}$ for the 0-20\% (scaled by a factor 100), the 20-40\% (scaled by a factor 10) and 40-80\% centrality classes compared to NLO pQCD predictions for the direct photon yield in pp collisions at the same energy, scaled by the number of binary nucleon collisions for each centrality class.}
\end{figure}

The inclusive photon spectra and double ratios of the PCM and PHOS analyses are combined as two independent measurements to obtain the error-weighted average. The uncertainties common to both measurements (trigger efficiency, centrality determination, etc.) are negligible in comparison to the uncorrelated, analysis-specific uncertainties. For each centrality selection the average double ratio $R_\gamma$ is used together with the averaged inclusive photon spectrum to obtain the final direct photon spectrum, according to Eq.~\ref{eq:subs}. For the 0--20\% centrality class and $\pT = \unit[2]{GeV}/c$, this results in type~A, B, and C systematic uncertainties of $\sigma_\mathrm{A}=2.5\%$, $\sigma_\mathrm{B}=2.3\%$, and $\sigma_\mathrm{C}=3.0\%$ for the combined double ratio and of  $\sigma_\mathrm{A}=20 \%$, $\sigma_\mathrm{B}=18 \%,$ $\sigma_\mathrm{C}=24 \%$ for the combined direct photon spectrum.

The combined PCM and PHOS double ratios $R_\gamma$ measured for three centrality classes are shown in Fig.~\ref{fig:rgamma}. A direct photon excess is observed for all centrality classes for $\pT \gtrsim \unit[4]{GeV}/c$, and also for $1 \lesssim \pT \lesssim \unit[4]{GeV}/c$ in the most central class. The measurements are compared with the expected $R_\gamma$ for the prompt photon contribution as calculated with next-to-leading-order (NLO) perturbative QCD calculations. The prompt photon expectations in Fig.~\ref{fig:rgamma} were determined as $1 + N_\mathrm{coll}  \gamma_\mathrm{pQCD} / \gamma_\mathrm{decay} $ where the number of binary nucleon-nucleon collisions ($N_\mathrm{coll} = 1210.8 \pm 132.5$, $438.4 \pm 42$, and $77.2 \pm 18$ for the 0--20\%,  20--40\%, and 40--80\% class, respectively) was calculated with a Monte Carlo Glauber code \cite{Abelev:2013qoq} using an inelastic nucleon-nucleon cross section of $\sigma_\mathrm{NN}^\mathrm{inel} = \unit[64 \pm 5]{mb}$. The decay photon spectra $\gamma_\mathrm{decay}$ were calculated as the product of the $(\gamma_\mathrm{decay}/\pi^0)|_\mathrm{MC}$ ratio from the decay photon calculation and the combined PHOS and PCM $\pi^0$ spectra. Three different direct photon calculations are shown, two based on JETPHOX (with different parton distribution functions) \cite{Klasen:2013mga}, and one from Refs.~\cite{Gordon:1993qc,Vog97a}. The band around the latter reflects the factorization, renormalization, and fragmentation scale uncertainty whereas the bands around the JETPHOX calculations also include the uncertainty of the parton distribution functions. In all three centrality classes, the excess agrees with the calculated prompt direct photon contributions at high $\pT \gtrsim \unit[5]{GeV}/c$. The contribution of prompt direct photons cannot be calculated straightforwardly for $\pT \lesssim \unit[2]{GeV}/c$; their contribution relative to the decay photons, however, is expected to be small. The excess of about 10--15\% for the 0--20\% centrality class in the range $0.9 \lesssim \pT \lesssim \unit[2.1]{GeV}/c$ indicates the presence of another source of direct photons in central collisions. The significance of the excess at each data point in this $\pT$ range in the 0--20\% centrality class is about $2\sigma$. Considering all data points in $0.9 \lesssim \pT \lesssim \unit[2.1]{GeV}/c$, the significance of the direct photon excess is about $2.6\sigma$ which is only slightly larger than the significance of the individual points due to the correlation of systematic uncertainties in $\pT$.

The resulting direct photon spectra are shown in Fig.~\ref{fig:spectra}. Arrows represent 90\% upper confidence limits. The same NLO pQCD calculations that were used in Fig.~\ref{fig:rgamma} are directly compared with the measured direct-photon spectra. In addition, the pQCD calculation used in the Pb-Pb direct photon prediction by Paquet et al.\ \cite{Paquet:2015lta} is shown as a dashed line in Fig.~\ref{fig:spectra}. This calculation was performed down to $\pT \approx \unit[1]{GeV}/c$ by using large scales $\mu$ ($> 2 \pt^\gamma)$ and rescaling the result so that it agrees with a calculation done with smaller scales at higher $\pT$. The systematic uncertainty of this calculation is estimated to be about 25\% for $\pT \gtrsim \unit[5]{GeV}/c$, growing to about 60\% at $\pT \approx \unit[1]{GeV}/c$. All calculations were scaled with the corresponding number of nucleon-nucleon collisions in the centrality class. Similar to $R_\gamma$, an agreement with these theoretical estimates of pQCD photon production in peripheral, mid-central, and central collisions for $\pT \gtrsim \unit[5]{GeV}/c$ is found. An agreement between $N_\mathrm{coll}$-scaled pQCD calculation and data for {\it isolated} direct photon yields was also found at higher $\pT$ ($> \unit[20]{GeV}/c$) by ATLAS \cite{Aad:2015lcb} and CMS \cite{Chatrchyan:2012vq}.

\begin{figure}[h]
\unitlength\textwidth
\centering
\includegraphics[width=0.7\linewidth]{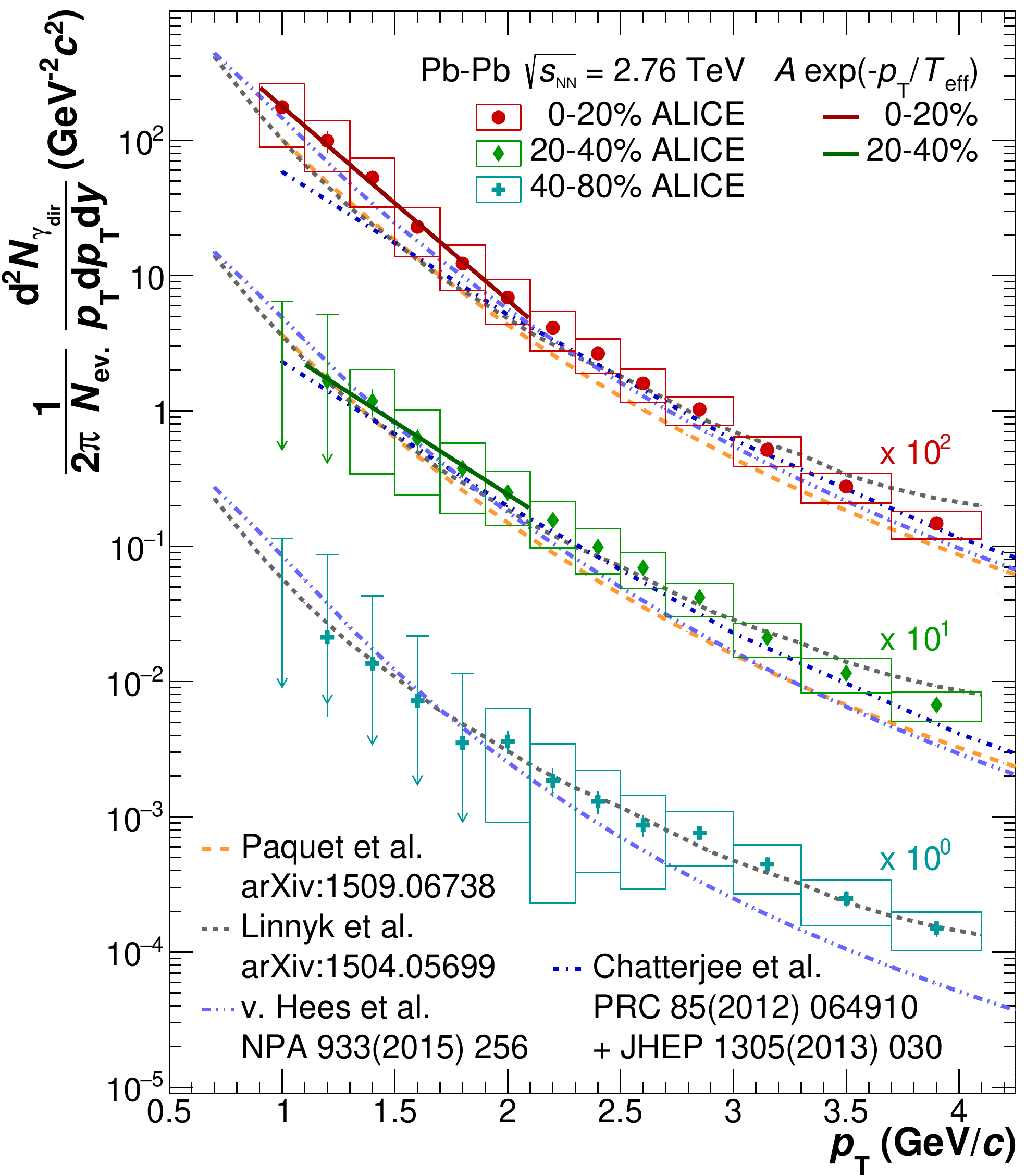}
\caption{
\label{fig:spectraAndThermal} 
(Color online) Comparison of model calculations from Refs.~\cite{vanHees:2014ida,Chatterjee:2012dn,Paquet:2015lta,Linnyk:2015tha} with the direct photon spectra in \PbPb\ collisions at $\snn=\unit[2.76]{TeV}$ for the 0--20\% (scaled by a factor 100), the 20--40\% (scaled by a factor 10) and 40--80\% centrality classes. All models include a contribution from pQCD photons. For the 0--20\% and 20--40\%  classes the fit with an exponential function is shown in addition.}
\end{figure}

In mid-central and more clearly in central collisions an excess of direct photons at low $\pT \lesssim \unit[4]{GeV}/c$ with respect to the pQCD photon predictions is observed, which might be related to the production of thermal photons. In models in which thermal photon production in the early phase dominates, the inverse slope parameter reflects an effective temperature averaged over the different temperatures during the space-time evolution of the medium. In order to extract the slope parameter, a $\pT$ region is selected where the contribution of prompt direct photons is small. The pQCD contribution from the calculation by Paquet et al.\ \cite{Paquet:2015lta}, shown as a dashed line in Fig.~\ref{fig:spectra}, is subtracted and the remaining excess yield is fit with an exponential function $\propto \exp(-\pT/T_\mathrm{eff})$. The extracted inverse slope parameter is $T_\mathrm{eff} = \SlopeCentral$ in the range $\FitrangeCentral$ for the 0--20\% class and $T_\mathrm{eff} = \SlopeSemiCentral$ in the range $\FitrangeSemiCentral$ for the 20--40\% class. Alternatively, to estimate the sensitivity to the pQCD photon contribution, the slope was extracted without the subtraction of pQCD photons. This yields inverse slopes of $T_\mathrm{eff}^\mathrm{no\,subtr} = \SlopeCentralNoSubtr$ for the 0--20\% class and $T_\mathrm{eff}^\mathrm{no\,subtr} = \SlopeSemiCentralNoSubtr$ for the 20--40\% class. The dominant contribution to the systematic uncertainty of the inverse slopes is due to the type~B uncertainties.

A significant contribution of blueshifted photons from the late stages of the collision evolution with high radial flow velocities has to be taken into account \cite{vanHees:2011vb,Shen:2013vja}. This makes the relation between the medium temperature and the inverse slope parameter less direct and a comparison to full direct photon calculations including the photons emitted during the QGP and hadron gas phase is necessary to extract the initial temperature. A comparison to state-of-the-art direct photon calculations is shown in Fig.~\ref{fig:spectraAndThermal}. All shown models assume the formation of a QGP. The hydrodynamic models, which fold the space-time evolution with photon production rates, use QGP rates from Ref.~\cite{Arnold:2001ms} and equations of state from lattice QCD. All models include the contribution from pQCD photons, however, different parameterizations are used. The model of van~Hees et al. \cite{vanHees:2014ida} is based on ideal hydrodynamics with initial flow (prior to thermalization) \cite{Seon:2014}. The photon production rates in the hadronic phase are based on a massive Yang-Mills description of gas of $\pi$, $K$, $\rho$, $K^*$, and $a_1$ mesons, along with additional production channels (including anti-/baryons) evaluated with the in-medium $\rho$ spectral function \cite{Turbide:2003si}. Bremsstrahlung from $\pi$--$\pi$ and $K$--$\bar K$ is also included \cite{Heffernan:2014mla}, in the calculation shown here together with $\pi$--$\rho$--$\omega$ channels recently described in Ref.~\cite{Holt:2015cda}. The space-time evolution starts at $\tau_0 = \unit[0.2]{fm}/c$ with temperatures $T_0 = 682$, $641$, $\unit[461]{MeV}$ for the 0--20\%, 20--40\%, and 40--80\% classes, respectively, at the center of the fireball. The calculation by Chatterjee et al.\ \cite{Chatterjee:2012dn,Helenius:2013bya} is based on an event-by-event (2+1D) longitudinally boost invariant ideal hydrodynamic model with fluctuating initial conditions. An earlier prediction with smooth initial conditions was presented in Ref.~\cite{Holopainen:2011pd}. Hadron gas rates are taken from the massive Yang-Mills approach of Ref.~\cite{Turbide:2003si}. Bremsstrahlung from hadron scattering is not included. The hydrodynamic evolution in the model of Chatterjee et al.\ starts at $\tau_0 = \unit[0.14]{fm}/c$ with an average temperature at the center of the fireball of $T_0 \approx \unit[740]{MeV}$ for the 0--20\% class and $T_0 \approx \unit[680]{MeV}$ for the 20--40\% class. The calculation by Paquet et al.\ \cite{Paquet:2015lta} uses event-by-event (2+1D) longitudinally boost invariant viscous hydrodynamics \cite{Ryu:2015vwa} with IP-Glasma initial conditions \cite{Schenke:2012wb}. Viscous corrections were applied to the photon production rates \cite{Dion:2011pp,Shen:2014nfa,Paquet:2015lta}. The same hadron gas rates as described above for the calculation by van Hees et al.\ are used. The hydrodynamic evolution starts at $\tau_0 = \unit[0.4]{fm}/c$ with an initial temperature (averaged over all volume elements with $T > \unit[145]{MeV}$) of $T_0 = \unit[385]{MeV}$ for the 0--20\% class and $T_0 = \unit[350]{MeV}$ for the 20--40\% class. The PHSD model prediction by Linnyk et al.\ \cite{Linnyk:2015tha} is based on an off-shell transport approach in which the full evolution of the collision is described microscopically. Bremsstrahlung from the scattering of hadrons is a significant photon source in this model. The comparison of the measured direct-photon spectra to the calculations in Fig.~\ref{fig:spectraAndThermal} indicates that the systematic uncertainties do not allow us to discriminate between the models.

\section{Conclusions}
The $\pt$ differential invariant yield of direct photons has been measured for the first time in \PbPb\ collisions at $\snn = \unit[2.76]{TeV}$ for transverse momenta $0.9< \pT < \unit[14]{GeV}/c$ and for three centrality classes: 0--20\%, 20--40\%, and 40--80\%. Two independent and consistent measurements (PCM, PHOS) have been averaged to obtain the final results. In all centrality classes, the spectra at high transverse momentum $\pT\gtrsim \unit[5]{GeV}/c$ follow the expectation from pQCD calculations of the direct photon yield in pp collisions at the same energy, scaled by the number of binary nucleon collisions. 
Within the sensitivity of the current measurement, no evidence for medium influence on direct photon production at high $\pT$ is observed. In the low $\pt$ region, $\pT \lesssim \unit[2]{GeV}/c$, no direct photon signal can be extracted in peripheral collisions, but in mid-central and central collisions an excess above the prompt photon contributions is observed. An inverse slope parameter of $T_\mathrm{eff} = \SlopeCentral$ is obtained for the 0--20\% most central collisions from an exponential function fit to the direct photon spectrum, after subtraction of the pQCD contribution, in the range $\FitrangeCentral$. Models which assume the formation of a QGP were found to agree with the measurements within uncertainties.

\newenvironment{acknowledgement}{\relax}{\relax}
\begin{acknowledgement}
\section*{Acknowledgements}
We would like to thank 
Rupa Chatterjee,
Olena Linnyk,
Jean-Fran\c{c}ois Paquet,
Ralf Rapp,
and Werner Vogelsang 
for providing calculations shown in this paper and for useful discussions.

The ALICE Collaboration would like to thank all its engineers and technicians for their invaluable contributions to the construction of the experiment and the CERN accelerator teams for the outstanding performance of the LHC complex.
The ALICE Collaboration gratefully acknowledges the resources and support provided by all Grid centres and the Worldwide LHC Computing Grid (WLCG) collaboration.
The ALICE Collaboration acknowledges the following funding agencies for their support in building and
running the ALICE detector:
State Committee of Science,  World Federation of Scientists (WFS)
and Swiss Fonds Kidagan, Armenia;
Conselho Nacional de Desenvolvimento Cient\'{\i}fico e Tecnol\'{o}gico (CNPq), Financiadora de Estudos e Projetos (FINEP),
Funda\c{c}\~{a}o de Amparo \`{a} Pesquisa do Estado de S\~{a}o Paulo (FAPESP);
National Natural Science Foundation of China (NSFC), the Chinese Ministry of Education (CMOE)
and the Ministry of Science and Technology of China (MSTC);
Ministry of Education and Youth of the Czech Republic;
Danish Natural Science Research Council, the Carlsberg Foundation and the Danish National Research Foundation;
The European Research Council under the European Community's Seventh Framework Programme;
Helsinki Institute of Physics and the Academy of Finland;
French CNRS-IN2P3, the `Region Pays de Loire', `Region Alsace', `Region Auvergne' and CEA, France;
German Bundesministerium fur Bildung, Wissenschaft, Forschung und Technologie (BMBF) and the Helmholtz Association;
General Secretariat for Research and Technology, Ministry of Development, Greece;
Hungarian Orszagos Tudomanyos Kutatasi Alappgrammok (OTKA) and National Office for Research and Technology (NKTH);
Department of Atomic Energy and Department of Science and Technology of the Government of India;
Istituto Nazionale di Fisica Nucleare (INFN) and Centro Fermi -
Museo Storico della Fisica e Centro Studi e Ricerche ``Enrico Fermi'', Italy;
MEXT Grant-in-Aid for Specially Promoted Research, Ja\-pan;
Joint Institute for Nuclear Research, Dubna;
National Research Foundation of Korea (NRF);
Consejo Nacional de Cienca y Tecnologia (CONACYT), Direccion General de Asuntos del Personal Academico(DGAPA), M\'{e}xico, Amerique Latine Formation academique - 
European Commission~(ALFA-EC) and the EPLANET Program~(European Particle Physics Latin American Network);
Stichting voor Fundamenteel Onderzoek der Materie (FOM) and the Nederlandse Organisatie voor Wetenschappelijk Onderzoek (NWO), Netherlands;
Research Council of Norway (NFR);
National Science Centre, Poland;
Ministry of National Education/Institute for Atomic Physics and National Council of Scientific Research in Higher Education~(CNCSI-UEFISCDI), Romania;
Ministry of Education and Science of Russian Federation, Russian
Academy of Sciences, Russian Federal Agency of Atomic Energy,
Russian Federal Agency for Science and Innovations and The Russian
Foundation for Basic Research;
Ministry of Education of Slovakia;
Department of Science and Technology, South Africa;
Centro de Investigaciones Energeticas, Medioambientales y Tecnologicas (CIEMAT), E-Infrastructure shared between Europe and Latin America (EELA), 
Ministerio de Econom\'{i}a y Competitividad (MINECO) of Spain, Xunta de Galicia (Conseller\'{\i}a de Educaci\'{o}n),
Centro de Aplicaciones Tecnológicas y Desarrollo Nuclear (CEA\-DEN), Cubaenerg\'{\i}a, Cuba, and IAEA (International Atomic Energy Agency);
Swedish Research Council (VR) and Knut $\&$ Alice Wallenberg
Foundation (KAW);
Ukraine Ministry of Education and Science;
United Kingdom Science and Technology Facilities Council (STFC);
The United States Department of Energy, the United States National
Science Foundation, the State of Texas, and the State of Ohio;
Ministry of Science, Education and Sports of Croatia and  Unity through Knowledge Fund, Croatia;
Council of Scientific and Industrial Research (CSIR), New Delhi, India;
Pontificia Universidad Cat\'{o}lica del Per\'{u}.
\end{acknowledgement}

\ifcernpp
\bibliographystyle{utphys} 
\else
\bibliographystyle{h-physrev}
\fi
\bibliography{master}

\newpage
\appendix
\section{The ALICE Collaboration}
\label{app:collab}



\begingroup
\small
\begin{flushleft}
J.~Adam\Irefn{org40}\And
D.~Adamov\'{a}\Irefn{org83}\And
M.M.~Aggarwal\Irefn{org87}\And
G.~Aglieri Rinella\Irefn{org36}\And
M.~Agnello\Irefn{org110}\And
N.~Agrawal\Irefn{org48}\And
Z.~Ahammed\Irefn{org132}\And
S.U.~Ahn\Irefn{org68}\And
S.~Aiola\Irefn{org136}\And
A.~Akindinov\Irefn{org58}
% ...
\renewcommand\labelenumi{\textsuperscript{\theenumi}~}

\section*{Affiliation notes}
\renewcommand\theenumi{\roman{enumi}}
\begin{Authlist}
\item \Adef{0}Deceased
\item \Adef{idp1747472}{Also at: Georgia State University, Atlanta, Georgia, United States}
\item \Adef{idp3783680}{Also at: M.V. Lomonosov Moscow State University, D.V. Skobeltsyn Institute of Nuclear, Physics, Moscow, Russia}
\end{Authlist}

\section*{Collaboration Institutes}
\renewcommand\theenumi{\arabic{enumi}~}
\begin{Authlist}

\item \Idef{org1}A.I. Alikhanyan National Science Laboratory (Yerevan Physics Institute) Foundation, Yerevan, Armenia
\item \Idef{org2}Benem\'{e}rita Universidad Aut\'{o}noma de Puebla, Puebla, Mexico
\item \Idef{org3}Bogolyubov Institute for Theoretical Physics, Kiev, Ukraine
\item \Idef{org4}Bose Institute, Department of Physics and Centre for Astroparticle Physics and Space Science (CAPSS), Kolkata, India
\item \Idef{org5}Budker Institute for Nuclear Physics, Novosibirsk, Russia
\item \Idef{org6}California Polytechnic State University, San Luis Obispo, California, United States
\item \Idef{org7}Central China Normal University, Wuhan, China
\item \Idef{org8}Centre de Calcul de l'IN2P3, Villeurbanne, France
\item \Idef{org9}Centro de Aplicaciones Tecnol\'{o}gicas y Desarrollo Nuclear (CEADEN), Havana, Cuba
\item \Idef{org10}Centro de Investigaciones Energ\'{e}ticas Medioambientales y Tecnol\'{o}gicas (CIEMAT), Madrid, Spain
\item \Idef{org11}Centro de Investigaci\'{o}n y de Estudios Avanzados (CINVESTAV), Mexico City and M\'{e}rida, Mexico
\item \Idef{org12}Centro Fermi - Museo Storico della Fisica e Centro Studi e Ricerche ``Enrico Fermi'', Rome, Italy
\item \Idef{org13}Chicago State University, Chicago, Illinois, USA
\item \Idef{org14}China Institute of Atomic Energy, Beijing, China
\item \Idef{org15}Commissariat \`{a} l'Energie Atomique, IRFU, Saclay, France
\item \Idef{org16}COMSATS Institute of Information Technology (CIIT), Islamabad, Pakistan
\item \Idef{org17}Departamento de F\'{\i}sica de Part\'{\i}culas and IGFAE, Universidad de Santiago de Compostela, Santiago de Compostela, Spain
\item \Idef{org18}Department of Physics and Technology, University of Bergen, Bergen, Norway
\item \Idef{org19}Department of Physics, Aligarh Muslim University, Aligarh, India
\item \Idef{org20}Department of Physics, Ohio State University, Columbus, Ohio, United States
\item \Idef{org21}Department of Physics, Sejong University, Seoul, South Korea
\item \Idef{org22}Department of Physics, University of Oslo, Oslo, Norway
\item \Idef{org23}Dipartimento di Elettrotecnica ed Elettronica del Politecnico, Bari, Italy
\item \Idef{org24}Dipartimento di Fisica dell'Universit\`{a} 'La Sapienza' and Sezione INFN Rome, Italy
\item \Idef{org25}Dipartimento di Fisica dell'Universit\`{a} and Sezione INFN, Cagliari, Italy
\item \Idef{org26}Dipartimento di Fisica dell'Universit\`{a} and Sezione INFN, Trieste, Italy
\item \Idef{org27}Dipartimento di Fisica dell'Universit\`{a} and Sezione INFN, Turin, Italy
\item \Idef{org28}Dipartimento di Fisica e Astronomia dell'Universit\`{a} and Sezione INFN, Bologna, Italy
\item \Idef{org29}Dipartimento di Fisica e Astronomia dell'Universit\`{a} and Sezione INFN, Catania, Italy
\item \Idef{org30}Dipartimento di Fisica e Astronomia dell'Universit\`{a} and Sezione INFN, Padova, Italy
\item \Idef{org31}Dipartimento di Fisica `E.R.~Caianiello' dell'Universit\`{a} and Gruppo Collegato INFN, Salerno, Italy
\item \Idef{org32}Dipartimento di Scienze e Innovazione Tecnologica dell'Universit\`{a} del  Piemonte Orientale and Gruppo Collegato INFN, Alessandria, Italy
\item \Idef{org33}Dipartimento Interateneo di Fisica `M.~Merlin' and Sezione INFN, Bari, Italy
\item \Idef{org34}Division of Experimental High Energy Physics, University of Lund, Lund, Sweden
\item \Idef{org35}Eberhard Karls Universit\"{a}t T\"{u}bingen, T\"{u}bingen, Germany
\item \Idef{org36}European Organization for Nuclear Research (CERN), Geneva, Switzerland
\item \Idef{org37}Excellence Cluster Universe, Technische Universit\"{a}t M\"{u}nchen, Munich, Germany
\item \Idef{org38}Faculty of Engineering, Bergen University College, Bergen, Norway
\item \Idef{org39}Faculty of Mathematics, Physics and Informatics, Comenius University, Bratislava, Slovakia
\item \Idef{org40}Faculty of Nuclear Sciences and Physical Engineering, Czech Technical University in Prague, Prague, Czech Republic
\item \Idef{org41}Faculty of Science, P.J.~\v{S}af\'{a}rik University, Ko\v{s}ice, Slovakia
\item \Idef{org42}Faculty of Technology, Buskerud and Vestfold University College, Vestfold, Norway
\item \Idef{org43}Frankfurt Institute for Advanced Studies, Johann Wolfgang Goethe-Universit\"{a}t Frankfurt, Frankfurt, Germany
\item \Idef{org44}Gangneung-Wonju National University, Gangneung, South Korea
\item \Idef{org45}Gauhati University, Department of Physics, Guwahati, India
\item \Idef{org46}Helsinki Institute of Physics (HIP), Helsinki, Finland
\item \Idef{org47}Hiroshima University, Hiroshima, Japan
\item \Idef{org48}Indian Institute of Technology Bombay (IIT), Mumbai, India
\item \Idef{org49}Indian Institute of Technology Indore, Indore (IITI), India
\item \Idef{org50}Inha University, Incheon, South Korea
\item \Idef{org51}Institut de Physique Nucl\'eaire d'Orsay (IPNO), Universit\'e Paris-Sud, CNRS-IN2P3, Orsay, France
\item \Idef{org52}Institut f\"{u}r Informatik, Johann Wolfgang Goethe-Universit\"{a}t Frankfurt, Frankfurt, Germany
\item \Idef{org53}Institut f\"{u}r Kernphysik, Johann Wolfgang Goethe-Universit\"{a}t Frankfurt, Frankfurt, Germany
\item \Idef{org54}Institut f\"{u}r Kernphysik, Westf\"{a}lische Wilhelms-Universit\"{a}t M\"{u}nster, M\"{u}nster, Germany
\item \Idef{org55}Institut Pluridisciplinaire Hubert Curien (IPHC), Universit\'{e} de Strasbourg, CNRS-IN2P3, Strasbourg, France
\item \Idef{org56}Institute for Nuclear Research, Academy of Sciences, Moscow, Russia
\item \Idef{org57}Institute for Subatomic Physics of Utrecht University, Utrecht, Netherlands
\item \Idef{org58}Institute for Theoretical and Experimental Physics, Moscow, Russia
\item \Idef{org59}Institute of Experimental Physics, Slovak Academy of Sciences, Ko\v{s}ice, Slovakia
\item \Idef{org60}Institute of Physics, Academy of Sciences of the Czech Republic, Prague, Czech Republic
\item \Idef{org61}Institute of Physics, Bhubaneswar, India
\item \Idef{org62}Institute of Space Science (ISS), Bucharest, Romania
\item \Idef{org63}Instituto de Ciencias Nucleares, Universidad Nacional Aut\'{o}noma de M\'{e}xico, Mexico City, Mexico
\item \Idef{org64}Instituto de F\'{\i}sica, Universidad Nacional Aut\'{o}noma de M\'{e}xico, Mexico City, Mexico
\item \Idef{org65}iThemba LABS, National Research Foundation, Somerset West, South Africa
\item \Idef{org66}Joint Institute for Nuclear Research (JINR), Dubna, Russia
\item \Idef{org67}Konkuk University, Seoul, South Korea
\item \Idef{org68}Korea Institute of Science and Technology Information, Daejeon, South Korea
\item \Idef{org69}KTO Karatay University, Konya, Turkey
\item \Idef{org70}Laboratoire de Physique Corpusculaire (LPC), Clermont Universit\'{e}, Universit\'{e} Blaise Pascal, CNRS--IN2P3, Clermont-Ferrand, France
\item \Idef{org71}Laboratoire de Physique Subatomique et de Cosmologie, Universit\'{e} Grenoble-Alpes, CNRS-IN2P3, Grenoble, France
\item \Idef{org72}Laboratori Nazionali di Frascati, INFN, Frascati, Italy
\item \Idef{org73}Laboratori Nazionali di Legnaro, INFN, Legnaro, Italy
\item \Idef{org74}Lawrence Berkeley National Laboratory, Berkeley, California, United States
\item \Idef{org75}Moscow Engineering Physics Institute, Moscow, Russia
\item \Idef{org76}Nagasaki Institute of Applied Science, Nagasaki, Japan
\item \Idef{org77}National Centre for Nuclear Studies, Warsaw, Poland
\item \Idef{org78}National Institute for Physics and Nuclear Engineering, Bucharest, Romania
\item \Idef{org79}National Institute of Science Education and Research, Bhubaneswar, India
\item \Idef{org80}Niels Bohr Institute, University of Copenhagen, Copenhagen, Denmark
\item \Idef{org81}Nikhef, Nationaal instituut voor subatomaire fysica, Amsterdam, Netherlands
\item \Idef{org82}Nuclear Physics Group, STFC Daresbury Laboratory, Daresbury, United Kingdom
\item \Idef{org83}Nuclear Physics Institute, Academy of Sciences of the Czech Republic, \v{R}e\v{z} u Prahy, Czech Republic
\item \Idef{org84}Oak Ridge National Laboratory, Oak Ridge, Tennessee, United States
\item \Idef{org85}Petersburg Nuclear Physics Institute, Gatchina, Russia
\item \Idef{org86}Physics Department, Creighton University, Omaha, Nebraska, United States
\item \Idef{org87}Physics Department, Panjab University, Chandigarh, India
\item \Idef{org88}Physics Department, University of Athens, Athens, Greece
\item \Idef{org89}Physics Department, University of Cape Town, Cape Town, South Africa
\item \Idef{org90}Physics Department, University of Jammu, Jammu, India
\item \Idef{org91}Physics Department, University of Rajasthan, Jaipur, India
\item \Idef{org92}Physik Department, Technische Universit\"{a}t M\"{u}nchen, Munich, Germany
\item \Idef{org93}Physikalisches Institut, Ruprecht-Karls-Universit\"{a}t Heidelberg, Heidelberg, Germany
\item \Idef{org94}Purdue University, West Lafayette, Indiana, United States
\item \Idef{org95}Pusan National University, Pusan, South Korea
\item \Idef{org96}Research Division and ExtreMe Matter Institute EMMI, GSI Helmholtzzentrum f\"ur Schwerionenforschung, Darmstadt, Germany
\item \Idef{org97}Rudjer Bo\v{s}kovi\'{c} Institute, Zagreb, Croatia
\item \Idef{org98}Russian Federal Nuclear Center (VNIIEF), Sarov, Russia
\item \Idef{org99}Russian Research Centre Kurchatov Institute, Moscow, Russia
\item \Idef{org100}Saha Institute of Nuclear Physics, Kolkata, India
\item \Idef{org101}School of Physics and Astronomy, University of Birmingham, Birmingham, United Kingdom
\item \Idef{org102}Secci\'{o}n F\'{\i}sica, Departamento de Ciencias, Pontificia Universidad Cat\'{o}lica del Per\'{u}, Lima, Peru
\item \Idef{org103}Sezione INFN, Bari, Italy
\item \Idef{org104}Sezione INFN, Bologna, Italy
\item \Idef{org105}Sezione INFN, Cagliari, Italy
\item \Idef{org106}Sezione INFN, Catania, Italy
\item \Idef{org107}Sezione INFN, Padova, Italy
\item \Idef{org108}Sezione INFN, Rome, Italy
\item \Idef{org109}Sezione INFN, Trieste, Italy
\item \Idef{org110}Sezione INFN, Turin, Italy
\item \Idef{org111}SSC IHEP of NRC Kurchatov institute, Protvino, Russia
\item \Idef{org112}Stefan Meyer Institut f\"{u}r Subatomare Physik (SMI), Vienna, Austria
\item \Idef{org113}SUBATECH, Ecole des Mines de Nantes, Universit\'{e} de Nantes, CNRS-IN2P3, Nantes, France
\item \Idef{org114}Suranaree University of Technology, Nakhon Ratchasima, Thailand
\item \Idef{org115}Technical University of Ko\v{s}ice, Ko\v{s}ice, Slovakia
\item \Idef{org116}Technical University of Split FESB, Split, Croatia
\item \Idef{org117}The Henryk Niewodniczanski Institute of Nuclear Physics, Polish Academy of Sciences, Cracow, Poland
\item \Idef{org118}The University of Texas at Austin, Physics Department, Austin, Texas, USA
\item \Idef{org119}Universidad Aut\'{o}noma de Sinaloa, Culiac\'{a}n, Mexico
\item \Idef{org120}Universidade de S\~{a}o Paulo (USP), S\~{a}o Paulo, Brazil
\item \Idef{org121}Universidade Estadual de Campinas (UNICAMP), Campinas, Brazil
\item \Idef{org122}University of Houston, Houston, Texas, United States
\item \Idef{org123}University of Jyv\"{a}skyl\"{a}, Jyv\"{a}skyl\"{a}, Finland
\item \Idef{org124}University of Liverpool, Liverpool, United Kingdom
\item \Idef{org125}University of Tennessee, Knoxville, Tennessee, United States
\item \Idef{org126}University of the Witwatersrand, Johannesburg, South Africa
\item \Idef{org127}University of Tokyo, Tokyo, Japan
\item \Idef{org128}University of Tsukuba, Tsukuba, Japan
\item \Idef{org129}University of Zagreb, Zagreb, Croatia
\item \Idef{org130}Universit\'{e} de Lyon, Universit\'{e} Lyon 1, CNRS/IN2P3, IPN-Lyon, Villeurbanne, France
\item \Idef{org131}V.~Fock Institute for Physics, St. Petersburg State University, St. Petersburg, Russia
\item \Idef{org132}Variable Energy Cyclotron Centre, Kolkata, India
\item \Idef{org133}Warsaw University of Technology, Warsaw, Poland
\item \Idef{org134}Wayne State University, Detroit, Michigan, United States
\item \Idef{org135}Wigner Research Centre for Physics, Hungarian Academy of Sciences, Budapest, Hungary
\item \Idef{org136}Yale University, New Haven, Connecticut, United States
\item \Idef{org137}Yonsei University, Seoul, South Korea
\item \Idef{org138}Zentrum f\"{u}r Technologietransfer und Telekommunikation (ZTT), Fachhochschule Worms, Worms, Germany
\end{Authlist}
\endgroup

\end{document}